\documentclass[%
 reprint,
 superscriptaddress,
 amsmath,amssymb,
 aps,
prl,
]{revtex4-2}

\usepackage{graphicx}
\usepackage{dcolumn}
\usepackage{bm}
\usepackage{xcolor}
\usepackage{hyperref}

\begin{document}

\preprint{APS/123-QED}

\newcommand{\LBL}{\affiliation{Nuclear Science Division, Lawrence Berkeley National Laboratory, Berkeley, CA 94720, USA}}
\newcommand{\Lund}{\affiliation{Department of Physics, Lund University, 22100 Lund, Sweden}}
\newcommand{\UCBerkeley}{\affiliation{University of California, Berkeley, CA 94720, USA}}
\newcommand{\Strasbourg}{\affiliation{Universit\'e de Strasbourg, CNRS, IPHC UMR 7178, 67037 Strasbourg, France}}
\newcommand{\LLNL}{\affiliation{Lawrence Livermore National Laboratory, Livermore, CA 94550, USA}}
\newcommand{\SJSU}{\affiliation{San Jos\'e State University, San Jose, CA 95192, USA}}
\newcommand{\TAMU}{\affiliation{Department of Chemistry, Texas A\&M University, College Station, TX 77843, USA}}
\newcommand{\CI}{\affiliation{Cyclotron Institute, Texas A\&M University, College Station, TX, 77843 USA}}
\newcommand{\ANL}{\affiliation{Argonne National Laboratory, Lemont, IL 60439, USA}}
\newcommand{\Zurich}{\affiliation{Department of Chemistry and Applied Biosciences, ETH Z\"urich, Zurich, CH}}
\newcommand{\PSI}{\affiliation{Nuclear Energy and Safety Division, Paul Scherrer Institute, Villigen PSI, CH}}
\newcommand{\OSU}{\affiliation{Oregon State Univesity, Corvallis, OR 97331, USA}}
\newcommand{\Liverpool}{\affiliation{University of Liverpool, Liverpool, UK}}
\newcommand{\Manchester}{\affiliation{University of Manchester, Manchester, UK}}
\newcommand{\ORNL}{\affiliation{Oak Ridge National Laboratory, Oak Ridge, TN 37831, USA}}

\title{Towards the Discovery of New Elements:\\Production of Livermorium ($Z$=116) with $^{50}$Ti}

\author{J.M. Gates}\email[Corresponding Author: ]{jmgates@lbl.gov}\LBL
\author{R. Orford}\LBL
\author{D. Rudolph}\Lund\LBL

\author{C. Appleton}\LBL
\author{B.M. Barrios}\SJSU
\author{J.Y. Benitez}\LBL
\author{M. Bordeau}\Strasbourg
\author{W.~Botha}\SJSU
\author{C.M.~Campbell}\LBL
\author{J. Chadderton}\Liverpool
\author{A.T. Chemey}\OSU
\author{R.M. Clark}\LBL
\author{H.L. Crawford}\LBL
\author{J.D.~Despotopulos}\LLNL
\author{O.~Dorvaux}\Strasbourg
\author{N.E. Esker}\SJSU
\author{P.~Fallon}\LBL
\author{C.M.~Folden III}\CI\TAMU
\author{B.J.P.~Gall}\Strasbourg
\author{F.H.~Garcia}\altaffiliation[Present affiliation: ]{Department of Chemistry, The University of British Columbia, 2036 Main Mall, Vancouver, British Columbia, V6T 1Z1, Canada}\LBL
\author{P. Golubev}\Lund\LBL
\author{J.A.~Gooding}\LBL\UCBerkeley
\author{M. Grebo}\LBL\UCBerkeley
\author{K.E.~Gregorich}\LLNL
\author{M.~Guerrero}\SJSU
\author{R.A.~Henderson}\LLNL
\author{R.-D.~Herzberg}\Liverpool
\author{Y. Hrabar}\Lund
\author{T.T.~King}\ORNL
\author{M. Kireeff~Covo}\LBL
\author{A.S. Kirkland}\CI\TAMU
\author{R.~Kr\"ucken}\LBL
\author{E. Leistenschneider}\LBL
\author{E.M.~Lykiardopoulou}\LBL
\author{M.~McCarthy}\LBL\UCBerkeley
\author{J.A. Mildon}\CI\TAMU
\author{C.~M\"uller-Gatermann}\ANL
\author{L. Phair}\LBL
\author{J.L. Pore}\LBL
\author{E.~Rice}\LBL\UCBerkeley
\author{K.P.~Rykaczewski}\ORNL
\author{B.N.~Sammis}\LLNL
\author{L.G.~Sarmiento}\Lund
\author{D.~Seweryniak}\ANL
\author{D.K.~Sharp}\Manchester
\author{A.~Sinjari}\LBL\UCBerkeley
\author{P.~Steinegger}\Zurich\PSI
\author{M.A. Stoyer}\LLNL\LBL
\author{J.M.~Szornel}\LBL
\author{K. Thomas}\LLNL
\author{D.S. Todd}\LBL
\author{P. Vo}\SJSU
\author{V. Watson}\LBL
\author{P.T. Wooddy}\LLNL
\date{\today}

\begin{abstract}
The $^{244}$Pu($^{50}$Ti,$xn$)$^{294-x}$Lv reaction was investigated at Lawrence Berkeley National Laboratory's 88-Inch Cyclotron facility. The experiment was aimed at the production of a superheavy element with $Z\ge 114$ by irradiating an actinide target with a beam heavier than $^{48}$Ca. Produced Lv ions were separated from the unwanted beam and nuclear reaction products using the Berkeley Gas-filled Separator and implanted into a newly commissioned focal plane detector system. Two decay chains were observed and assigned to the decay of $^{290}$Lv. The production cross section was measured to be $\sigma_{\rm prod}=0.44(^{+58}_{-28})$~pb at a center-of-target center-of-mass energy of 220(3)~MeV. This represents the first published measurement of the production of a superheavy element near the `Island-of-Stability', with a beam of $^{50}$Ti and is an essential precursor in the pursuit of searching for new elements beyond $Z=118$.
\end{abstract}

\maketitle


The production of SuperHeavy Elements (SHE, $Z> 103$), and the investigation of their nuclear properties, stands as an important frontier in modern nuclear physics, pushing the boundaries of our understanding of the fundamental constituents of matter \cite{NSAC2023}. The existence of SHE was first theorized in the 1950's as the result of stabilization of very heavy ($A\approx300$), neutron-rich ($N\approx184$) nuclei due to the presence of closed nuclear shells \cite{Sch1957,Mye1966,Sob1966,Nil1969}. Today, the concept of an `Island of Stability' remains an intriguing topic \cite{Smi2024}, with its exact position and extent on the Segr\'e chart continuing to be a subject of active pursuit both in theoretical and experimental nuclear physics \cite{Cwi1996, Ben1999, Kru2000,Cwi2005,Zha2005,Sob2007,Egi2020,Sam2021}.

Over the decades, SHE from $Z=104-118$ were discovered using different types of nuclear reactions: first by impinging light ions on actinide targets in so-called `hot'-fusion reactions ($Z\leq 106$) \cite{Hof2015}, and then by using transition metal beams (e.g., $^{50}$Ti, $^{51}$V, $^{54}$Cr, $^{62,64}$Ni, $^{70}$Zn) on targets of Pb or Bi, in so-called `cold'-fusion reactions ($Z\leq 113$). The production of SHE from both of these reaction mechanisms showed similar properties – quickly decreasing cross sections with increasing $Z$ of the compound nucleus. The heaviest element produced with one of these reactions was Nh ($Z=113$), using the $^{209}$Bi($^{70}$Zn,$n$) reaction. At a cross section of just $\sigma_{\rm prod}=22(^{+20}_{-13})$~fb \cite{Mor2004,Mor2012}, only three decays of $^{278}$Nh were observed in over 500 days of beamtime, seeming to mark the end of new SHE production. Fortunately, a major breakthrough was underway with the production of SHE by irradiating actinide targets from $^{238}$U ($Z=92$) to $^{249}$Cf ($Z=98$) with beams of $^{48}$Ca ($Z=20$) \cite{Oga2017}. Between 2000 and 2016, five new elements were added to the periodic table \cite{Ohr2016} and over 50 isotopes with $Z=104-118$ were discovered \cite{NPA944}. Since many of these are located near the `Island of Stability', these discoveries have provided crucial insights into the chemistry and physics of SHE \cite{EPJ2016}. One of the key focuses of the field is now on the production of new SHE.

\begin{figure}[ht]
\vspace*{-6mm}
\hspace*{-1mm}
\includegraphics[width=0.5\textwidth]{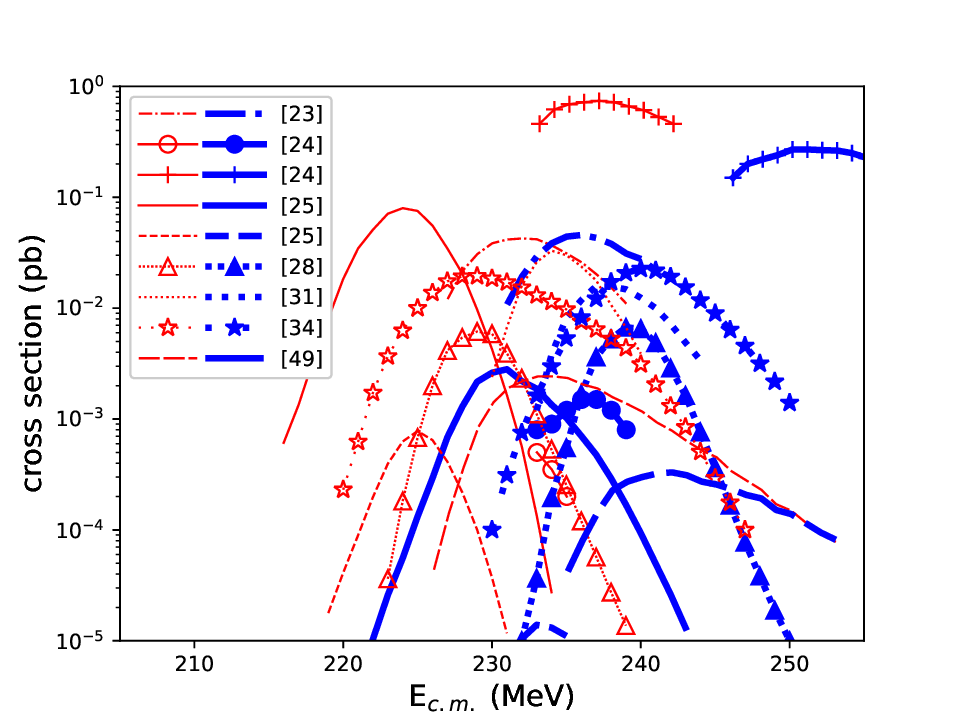}
\caption{\label{fig:cross_sections} Theoretical predictions of cross sections for the production of element $Z=120$ from the $3n$ (thin, red lines) and $4n$ (thick, blue lines) exit channels of the $^{50}$Ti+$^{249}$Cf reaction taken from \cite{Nas2011,Zag2008,Siw2012,Liu2013,Zhu2014,Ada2009,Zha2024}.}
\end{figure}

Presently, oganesson (Og, $Z=118$) marks the limit for the production of SHE using $^{48}$Ca beams. To attempt production of elements with $Z=119$ or $120$ using $^{48}$Ca, targets of Es ($Z=99$) or Fm ($Z=100$) would be required. Unfortunately, neither of these elements can be produced in sufficient quantities to produce a suitable target \cite{Rob2023}. Therefore, a new reaction approach is required. In this pursuit, there have been numerous theoretical studies aimed at predicting the production rate of new elements using actinide targets and ion beams heavier than $^{48}$Ca, such as $^{50}$Ti, $^{51}$V, or $^{54}$Cr \cite{Zag2008,Ada2009,Nas2011,Wan2011,Liu2011,Siw2012,Wan2012,Kuz2012,Liu2013,Jia2013,Zha2013,Zhu2014,Zag2015,Ada2020, Niu2021,Li2022,Li2022b,Zhu2023,Li2023}. Most of these models are able to reproduce the known excitation functions for the production of SHE with $^{48}$Ca beams on actinide targets reasonably well. They also largely agree that reactions with $^{50}$Ti have the highest cross sections for the production of elements with $Z=119$ and $120$ when compared to the other beam species. But the similarities end there. As shown in Fig.~\ref{fig:cross_sections}, the predicted cross sections for the $^{50}$Ti+$^{249}$Cf reaction span more than three orders of magnitude. Further, proposed ion-beam energies for maximum production differ by tens of MeV. Notably, these predictions are highly sensitive to the mass models used in the calculations \cite{Zag2008,Ada2009} and there are no mass measurements in the region with which to anchor the mass models. The disagreements within theoretical cross sections are currently hindering experimental efforts to produce new elements: The expected low cross sections imply that only one event every few weeks or months could be detected under ideal experimental settings. Another complication comes in planning these experiments, where choosing the correct excitation energy of the compound nucleus that corresponds to the maximum cross section is absolutely critical. If experimental settings are off by only a few MeV, the production rate may decrease dramatically.

Several experimental campaigns have attempted to make new elements with $Z=119$, $120$, and $122$ using the reactions $^{64}$Ni+$^{238}$U \cite{Hof2008}, $^{58}$Fe+$^{244}$Pu \cite{Oga2009}, $^{50}$Ti+$^{249}$Bk \cite{Khu2020}, $^{50}$Ti+$^{249}$Cf \cite{Khu2020}, and $^{70}$Zn+$^{238}$U \cite{Hof2002}. All have been unsuccessful to date, reaching one-event cross section limits of 0.09, 0.4, 0.065, 0.2, and 7.2~pb, respectively. Notably, these published upper-limit values are not able to sufficiently constrain theoretical predictions. Recently, a press release from the Joint Institute for Nuclear Research (JINR) claimed the production of the new isotope $^{288}$Lv in the reaction $^{54}$Cr+$^{238}$U \cite{PR2023}. However, no publication is presently available regarding the observed event(s), the measured cross section, or the utilized experimental setup. Additionally, an earlier publication reports on the possible production of element 120 using the reaction $^{54}$Cr+$^{248}$Cm \cite{Hof2016}. However, element 120 is still regarded as undiscovered, as other members of that collaboration attribute the same decay chain to a sequence of random events \cite{Hess2017}.

Given the status of new element searches, it is clearly important to test these production mechanisms for elements where the cross sections are predicted to be more accessible. For this reason, we chose to investigate the production of Lv ($Z=116$) using the $^{50}$Ti+$^{244}$Pu reaction. Several groups have published theoretical excitation functions or cross section predictions for both this reaction and reactions with $^{50}$Ti beams to make elements with $Z\ge119$ \cite{Zag2008,Kuz2012,Ada2020,Cap2023}. For example, the authors of Ref.~\cite{Zag2008} predict that $^{290}$Lv can be produced at a cross section of $\approx0.2$~pb at an excitation energy of $\approx45$~MeV, whereas the authors of Ref.~\cite{Ada2020} report a maximum cross section of $\approx0.1$~pb at an excitation energy of $\approx39$~MeV. Ref.~\cite{Kuz2012} contains two predictions created with different mass models, both of which give a cross section of $\approx0.05$~pb at an excitation energy of $\approx40$~MeV. A further calculation indicates that the cross section is between 0.12 and 0.86~pb \cite{Cap2023}. Hence, measuring the cross section of this reaction experimentally would give an important benchmark for constraining theoretical predictions.

Here we report on the first results from the $^{244}$Pu($^{50}$Ti,$xn$)$^{294-x}$Lv experiment using the Berkeley Gas-filled Separator (BGS) \cite{Gre2013} at Lawrence Berkeley National Laboratory's (LBNL) 88-Inch Cyclotron facility. This marks the beginning of a new era of SHE production and research utilizing beams beyond $^{48}$Ca. 

Isotopically enriched $^{50}$Ti ($\ge$90$\%$) was acquired as $^{50}$TiO$_2$ and reduced to its metallic form at Argonne National Laboratory (ANL). During four experimental campaigns, the metallic $^{50}$Ti was then used to produce a $^{50}$Ti$^{12+}$ beam from the Versatile ECR for NUclear Science (VENUS) ion source \cite{Lei2003,Lei2005} using a newly-developed induction oven \cite{Tod2024}. The average beam intensity out of VENUS was $\approx100~e\mu$A. This beam was accelerated to energies of 282(3)~MeV using the LBNL 88-Inch Cyclotron. The beam energy was measured at the start of each campaign by non-destructively measuring the time-of-flight of individual beam pulses between two fast-current transformers separated by 3.563(5)~m along a neighboring beamline \cite{Kir2014}. The values obtained from this process were used to ensure that the relative energy did not change between the different experimental campaigns. The average $^{50}$Ti beam intensity was $\approx6\times 10^{12}$ ions per second at the exit of the cyclotron. After acceleration, the beam passed through a differential pumping section that isolates the vacuum of the cyclotron from the 0.45-Torr He fill gas within the BGS. There are several collimators within the differential pumping section that may slightly reduce the beam intensity on target as compared to that at the exit of the cyclotron. 

The beam then impinged on the target composed of four arc-shaped segments forming a rotating target wheel with a diameter of 12.2 cm. A fast-acting beam chopper, installed before the beam is inflected into the cyclotron, can interrupt the beam in case of system failures, protecting the target \cite{Kir2024}. Each target segment consisted of a 2.1(1)-$\mu$m thick $^{nat}$Ti backing foil onto which $^{244}$Pu had been electrodeposited. The electrodeposition was performed at Lawrence Livermore National Laboratory (LLNL). The targets were oriented in the beamline such that the beam first passed through the Ti foil before entering the $^{244}$Pu layer. Prior to beam irradiation, the target foils were measured to have an average target thicknesses of 0.435(40)~mg/cm$^2$ through $\gamma$-ray analysis of the decay of the short-lived $^{240m}$Np, which is part of the decay path originating from $^{244}$Pu. Note that some target material is sputtered during irradiation and that these targets were also previously irradiated for ten days during a $^{244}$Pu($^{48}$Ca,$xn$) experiment. They may thus be thinner than when initially produced.

To allow for cross-section calculations, two silicon pin-diode detectors were positioned at angles of $\pm 27.2(1)^\circ$ directly downstream of the target. These detectors monitor the integral of beam intensity times target thickness through the detection of Rutherford-scattered beam particles. The 
fraction of 4$\pi$ subtended by each detector is $\Omega=2.1(2)\times10^{-4}$.

Energy losses of the beam in the targets were assessed with SRIM2013 \cite{srim2010}. The beam is estimated to have lost 15(1)~MeV passing through the backing foil and an additional 
$3$ to $5$~MeV passing through the $^{244}$Pu target layer, depending on the target thickness of the segment. This yields an average center-of-target center-of-mass frame energy of 220(3)~MeV, which corresponds to an average compound-nucleus excitation energy of 41(2) MeV according to the Thomas-Fermi mass tables \cite{Mye1996}.

The targets were irradiated for a total of 22.1 days. During these measurements, the recoiling evaporation residues (EVRs) were separated from the beam and unwanted nuclear reaction products in the BGS \cite{Gre2013} based on their differing magnetic rigidities ($B\rho$) in 0.45-Torr He. The BGS was initially set to bend reaction products with $B\rho=2.19$~T$\cdot$m to its focal plane. This was increased to $B\rho$ = 2.24~T$\cdot$m for the last $\approx3.1$ days. The efficiency for transporting Lv EVRs through the BGS was estimated to be 70(7)$\%$ using the simulation code in \cite{Gre2013}. For the efficiency simulations, it was assumed that the BGS $B\rho$ was tuned such that Lv EVRs were centered in the focal-plane detector.

\begin{figure*}[tbh]
\includegraphics[width=\textwidth]{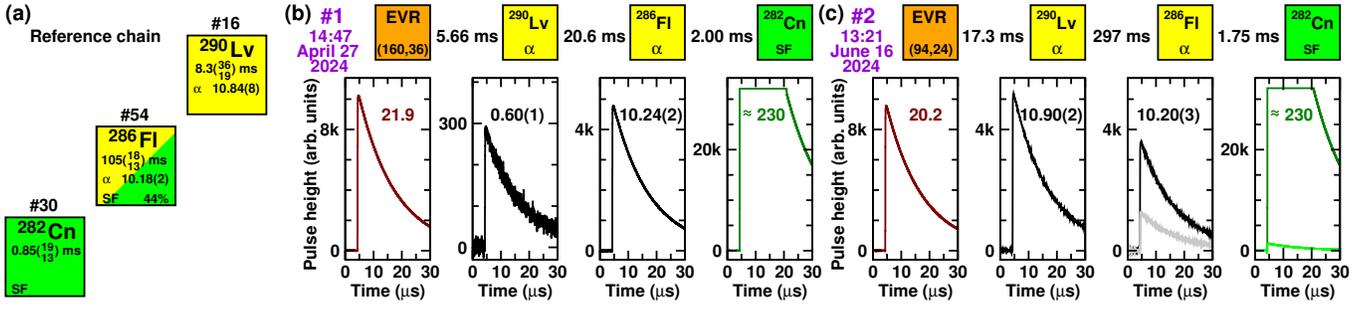}
\caption{\label{fig:chain} (a) Reference $\alpha$-decay chain of $^{290}$Lv \cite{SuppMat}. Lifetimes, $\alpha$-particle energies, and branching ratios are based on published data of decay events associated with $^{290}$Lv, $^{286}$Fl, and $^{282}$Cn \cite{Oga2004,Oga2006,Oga2004b,Oga2004c,Sta2009,Ell2010,Oga2012,Sam2021,Oga2022}. The number of events previously observed for each isotope is signified by \#$N$ above each isotope. (b) Waveforms of preamplifier pulses of the decay chain {\color{violet}\#1} assigned to $^{290}$Lv. Numbers in the panels are calibrated {\em detected} energies in MeV. Correlation times are given between recoil implantation (orange), $\alpha$ decays (yellow), and fission (green). The decay chain was observed in pixel (160,36). (c) Same as (b) but for decay chain {\color{violet}\#2} assigned to $^{290}$Lv. The decay chain was observed in pixel (94,24). The waveforms in lighter colors in the two rightmost graphs were registered in the neighboring pixel (93,24).
}
\end{figure*}

At the BGS focal plane, the EVRs were implanted into the SuperHeavy RECoil (SHREC) detector provided by Lund University \cite{Gol2024}. SHREC, and its read-out system were previously commissioned at the BGS focal plane using $^{254}$No EVRs produced in the $^{208}$Pb($^{48}$Ca,$2n$) reaction \cite{Orf2024} and $^{288-289}$Fl EVRs produced in the $^{244}$Pu($^{48}$Ca,3-4$n$) reaction. In brief, SHREC has an implantation detector that is situated perpendicular to the path of the beam. This detector is comprised of three side-by-side double-sided silicon-strip detectors (DSSDs). Each DSSD has an active area of $58.5\,\mathrm{mm}\times 58.5\,\mathrm{mm}$ and is subdivided into 58 strips on both the front side (junction) and the rear side (ohmic). On the front side of the detector, the 174 strips denote position in the horizontal direction. On the back sides of the detector, the 58 strips were wire-bonded across all three DSSDs, yielding 58 strips denoting vertical position. Directly downstream of the implantation detector is an identical set of three DSSDs that serve to veto signals from light, high-energy, charged particles. These particles pass through the 300 $\mu$m implantation detector, depositing only a portion of their energy in the implantation and veto detectors. They may thus mimic escape- and $\alpha$-like events. Upstream of the implantation detector is a `tunnel' of eight DSSDs which can catch the remaining energy fraction of $\alpha$ particles that escape from the face of the implantation detector. The geometric efficiency of SHREC for detecting a full-energy $\alpha$-particle in the implantation detector is $\gtrsim50$\%. Depending on the implantation profile, reconstructed $\alpha$ decays that split their energy deposition between the implantation detector and the upstream detectors increase the efficiency to $75$-$80\%$ \cite{Gol2024,Orf2024}.

Signals from all DSSDs were processed with compact charge-sensitive preamplifiers \cite{gol2013} and sent to ten 64-channel CAEN VX2740 digitizers (16 bit, up to 125~MS/s). Each digitizer channel self-triggered above an energy threshold of $\lesssim 200$~keV. Signals were processed using the Digital Pulse Processing Pulse Height Analysis (DPP-PHA) firmware controlled through the CoMPASS software from CAEN \cite{compass}. Waveforms (30-$\mu$s long), timestamps, detector strip identifiers, and uncalibrated `energies' from an online trapezoidal filter were recorded for all events. \cite{Gol2024,Orf2024}. Energy calibrations were performed for SHREC before and after each experiment using $\alpha$ sources consisting of $^{148}$Gd, $^{239}$Pu, $^{241}$Am, and $^{244}$Cm, as well as a $^{207}$Bi conversion-electron source. This calibration technique was previously optimized using $\alpha$-decay lines of $^{254}$No and $^{250}$Fm \cite{Orf2024}.

The expected reaction products of this experiment were from the $3n$ and $4n$ exit channels, $^{291}$Lv and $^{290}$Lv, respectively. The decay properties of both isotopes and their daughters have previously been published through their production both directly and indirectly in the $^{249}$Cf($^{48}$Ca,$3n$) \cite{Oga2004,Oga2006}, $^{245}$Cm($^{48}$Ca,$2$-$3n$) \cite{Oga2004,Oga2004b,Oga2006}, $^{244}$Pu($^{48}$Ca,$5n$) \cite{Oga2004} and $^{242}$Pu($^{48}$Ca,$3$-$4n$) \cite{Oga2004c,Sta2009,Ell2010,Sam2021,Oga2022} reactions. A discussion of the search parameters for decay chains originating from $^{291}$Lv is included in the supplemental material~\cite{SuppMat}. No decay chains were observed that fit the known decay properties of $^{291}$Lv and its daughters.

\begin{figure}[b]
\includegraphics[width=0.85\linewidth]{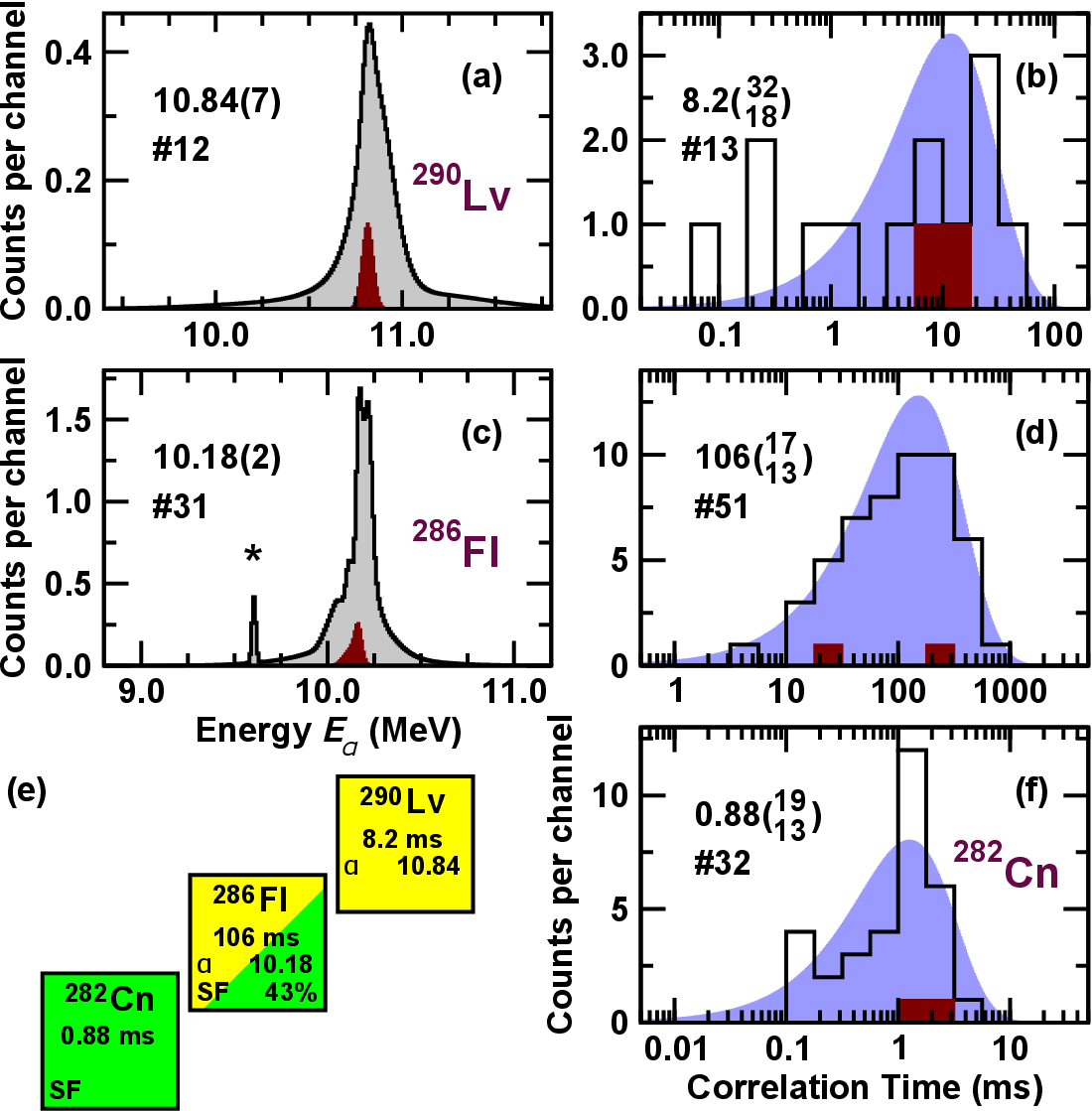}
\caption{\label{fig:Fig3-Curves}Compilation of information on the decays of $^{290}$Lv, $^{286}$Fl, and $^{282}$Cn \cite{Oga2004,Oga2006,Oga2004b,Oga2004c,Sta2009,Ell2010,Oga2012,Sam2021,Oga2022}. Panels (a) and (c) provide experimental decay-energy spectra from events associated with the decay steps $^{290}\mathrm{Lv}\rightarrow^{286}$Fl and $^{286}\mathrm{Fl}\rightarrow^{282}$Cn, respectively. For a single entry, a Gaussian with integral one and a width compliant with its measured uncertainty was added into the spectra. The numbers in the top left of these panels are the $\alpha$-decay energies, in MeV, extracted from the histogram mean in the intervals [10.0,11.7] and [9.9,10.5]~MeV, respectively. The right column [(b), (d), (f)] shows the correlation times of the decays along the decay chain starting with $^{290}$Lv. Experimental data points are comprised in the histograms (black lines). The shaded areas (blue) provide correlation-time distributions expected for the corresponding half life, $T_{1/2}$ in ms, which are given in the top left corner of each panel. For all panels, the number after the hashtag, $\#$, indicates the number of available data points. Entries marked in dark red correspond to the events associated with the observation of $^{290}$Lv in this work. The 9.6 MeV peak marked with an * in (c) was explained in detail in \cite{Sam2021}. Panel (e) shows the revised aggregated information of the $^{290}$Lv decay chain including the events from this work (cf.~Fig.~\ref{fig:chain}) \cite{SuppMat}.}
\end{figure}

Data from the published decay chain of $^{290}$Lv are summarized in Fig.~\ref{fig:chain}(a), Fig.~\ref{fig:Fig3-Curves}, and Ref.~\cite{SuppMat}. Potential decay chains originating from $^{290}$Lv were identified using correlations that required observing the implantation of an EVR (10$<E($MeV$)<$30) followed by the decay of at least one full-energy $\alpha$ (either $^{290}$Lv or $^{286}$Fl, [9.75$<E($MeV$)<$11.25]) followed by a spontaneous fission (SF) event ($E>120$~MeV). All three events must occur within the same $(x,y)$ pixel of the implantation detector and the SF must be within one second of the EVR. The efficiency for detecting a decay chain originating from $^{290}$Lv under these conditions is $\approx95\%$, based on Monte Carlo simulations of decay chains with branching ratios shown in Fig.~\ref{fig:chain}(a). The number of expected decay chains arising from correlations of random background events was calculated for each pixel individually, based on the rate of EVR-, $\alpha$-, and SF-like events in that pixel, then summed across the entire detector. The median rate of EVR-, $\alpha$-, and SF-like events within the above energy ranges was $1.2\times10^{-4}$, $1.4\times10^{-5}$, $5.1\times10^{-8}$~Hz/pixel, respectively. The probability for random background events to form a chain that would be detected using these search conditions is $1.7\times10^{-6}$.

Two decay chains were observed that met the criteria above. They are shown in Fig.~\ref{fig:chain}(b) and (c), including baseline-corrected waveforms of all constituent events. The first decay chain consisted of a 21.9-MeV EVR-like event followed 5.66 ms later by a 0.60(1)-MeV escape-like event in the same pixel. An $\alpha$-like event was observed 20.6~ms after the escape. The detected energy was $E=10.24(2)$~MeV, which includes the $\alpha$-particle energy and the energy from the recoiling daughter nucleus. Following procedures outlined in Ref.~\cite{Sam2023b}, the $\alpha$-particle energy was calculated to be $E_\alpha=10.16(2)$~MeV. The $\alpha$ energy and lifetime are consistent with the known decay properties of $^{286}$Fl and was assigned accordingly. Based on its observed lifetime and position in the decay chain, the 0.60(1)-MeV escape-like event was assigned to an $\alpha$-decay of $^{290}$Lv where the $\alpha$-particle escaped out of the front of the implantation detector and did not impact one of the upstream detectors. Thus, only a fraction of its decay energy was recorded (see, e.g., the spectra in Fig.~2(a) in the Supplemental Material of Ref.~\cite{Sam2023b}). The rate of escape-like events ($0.2<E($MeV$)<6.0$) was $7.8\times10^{-3}$~Hz per pixel. The probability for observing a random escape-like event in the $26$~ms between the EVR and the first observed full-energy $\alpha$ decay in the chain is $2.0\times10^{-4}$. The observed decay assigned to $^{286}$Fl was followed just 2.00~ms later by an $\approx230$~MeV SF-like event. The approximate energy of the SF-like event was determined by constructing an unsaturated waveform from the unsaturated portions of the recorded waveform using benchmarked pole-zero corrections \cite{Gol2024} and then extracting the pulse height using a trapezoidal energy filter. The lifetimes, decay modes, and decay energies of the events above are fully consistent with a decay chain consisting of a $^{290}$Lv EVR implanting into SHREC, followed by the $^{290}$Lv $\alpha$ escaping the front of SHREC, a full-energy $\alpha$-decay of $^{286}$Fl, and terminating with the SF of $^{282}$Cn [cf.~Fig.~\ref{fig:chain}(a) and Fig.~\ref{fig:Fig3-Curves}].

The second decay chain [Fig.~\ref{fig:chain}(c)] consisted of a 20.2 MeV EVR followed 17.3 ms later by a recoil-corrected $E_{\alpha}=$10.81(3)~MeV full-energy $\alpha$. A second full-energy $\alpha$ with $E_{\alpha}=10.12(4)$~MeV was detected 297~ms later. The decay chain was terminated by an $\approx$230~MeV SF-like event 1.75~ms after the second $\alpha$-particle. Based on the energies, lifetimes, and decay modes, this series of events was assigned to a decay chain consisting of an implanted $^{290}$Lv EVR followed by $\alpha$ decays of $^{290}$Lv and $^{286}$Fl, and terminating with SF of $^{282}$Cn. The probability of observing two chains composed of random background events based on the rates discussed above was $1.4\times10^{-12}$.

The first $^{290}$Lv decay chain was observed in pixel $(x,y)=(160,36)$ when the BGS $B\rho$ was set to 2.19 T$\cdot$m. The measured dispersion of the BGS is $16(2)~\textrm{mm}/\%(B\rho)$ \cite{Orf2024}, corresponding to a measured $B\rho=2.29(4)$~T$\cdot$m. The second event was observed in pixel $(x,y)=(94,24)$ and was detected after the BGS $B\rho$ was increased. The second event occurred at a $B\rho$=2.25(4)~T$\cdot$m, giving an average $B\rho$=2.27(3)~T$\cdot$m. 

The cross section for two events, derived from the observed number of Rutherford-scattered particles, is $\sigma_{\rm prod}=0.44(^{+58}_{-28})$~pb at the $68\%$ confidence level \cite{Sch1984,Sch2000}. The error represents statistical (counting) errors only. In addition to the cited statistical errors, there is a systematic error on the cross section. The factors contributing to systematic errors in a BGS experiment result from uncertainties in: (i) the efficiency for transporting an EVR from the target to SHREC, (ii) the angle of the Rutherford detectors and the Rutherford scattering cross sections, (iii) the solid angle subtended by the collimators upstream of the Rutherford detectors, (iv) the attenuation factors of the screens between the target and Rutherford detectors, and (v) the energy from the cyclotron. These five factors and how their associated errors were determined are discussed in detail in \cite{Gre2005} for experiments at the BGS and are expected to result in an additional $12\%$ systematic uncertainty in the measured cross sections. In cases where the reaction is run in the BGS for the first time and the $B\rho$ through the BGS is unknown, there is an additional uncertainty in detection efficiency.

The two-event cross section reported in this work is higher than theoretical predictions of Ref. \cite{Ada2020,Zag2008,Kuz2012}, and all three references can be excluded at the $68\%$ confidence level at the experimental excitation energy of 41(2)~MeV. The cross section 
is consistent with the theoretical prediction from Ref. \cite{Cap2023}, when theoretical uncertainties are included. The observation of the two events at an excitation energy of 41(2)~MeV is also consistent with the proposed optimal excitation energies in Refs.~\cite{Ada2020,Kuz2012}, although lower than that from \cite{Zag2008}.

The $4n$ reaction between $^{48}$Ca and $^{244}$Pu has been investigated previously and has been observed to have a cross section between $\sigma_{\rm prod}=5.3(^{+36}_{-21})$~pb \cite{Oga2004b,Sam2023b} and $\sigma_{\rm prod}=9.8(^{+39}_{-31})$~pb \cite{Gat2011}. These values are $\approx10$-$20$ times larger than the cross section reported in this work between $^{50}$Ti and $^{244}$Pu with the same exit channel. This indicates that the cross section for the production of element 120 with $^{50}$Ti beams could be $\approx25$-$50$~fb, based on the known $^{48}$Ca($^{249}$Cf,3$n$) cross section of $\sigma_{\rm prod}=0.5(^{+16}_{\;-3})$~pb \cite{Oga2006}, demonstrating that a substantial -- but seemingly manageable -- reduction in production cross sections has to be expected in the push towards discovering higher-$Z$ elements with beams beyond $^{48}$Ca. Production of SHE with $^{48}$Ca beams has been well-studied and excitation functions have been mapped out for $2$-$5n$ exit channels in many cases. Comparison of this systematic data to a single point on the $^{244}$Pu($^{50}$Ti,$4n$) excitation function is of high interest, although further investigations of these reactions with beams beyond $^{48}$Ca are crucial for constraining theoretical cross section calculations. Uncertainties in these predictions can only be reduced with the availability of additional experimental data across a wider range of excitation energies and beam/target combinations.

In summary, at the LBNL 88-Inch Cyclotron facility, a $^{244}$Pu target was irradiated with a high-intensity beam of $^{50}$Ti. Two decay chains were observed and assigned to the decay of $^{290}$Lv with a production cross section of $\sigma_{\rm prod}=0.44(^{+58}_{-28})$~pb at a center-of-target excitation energy of 41(2)~MeV. This is the first reported production of a SHE near the predicted `Island-of-Stability' with a beam other than $^{48}$Ca. While the cross section observed here does reflect the expected decrease in SHE production when moving to heavier beams, the success of this measurement validates that discoveries of new SHE are indeed within experimental reach.

\section{\label{sec:acknowledgements}Acknowledgements}
We gratefully acknowledge the operations staff of the 88-Inch Cyclotron for providing the intense beams of $^{50}$Ti and stable operating conditions. This work was supported in part by the U.S.~Department of Energy, Office of Science, Office of Nuclear Physics under contract numbers DE-AC02-05CH11231 (LBNL), DE-AC02-06CH11357 (ANL) and DE-FG02-93ER40773 (TAMU); the Swedish Knut and Alice Wallenberg Foundation (KAW 2015.0021), the Wenner-Gren Foundations (SSv2020-0003), the Carl Trygger Foundation (CTS 20:1146), and the Royal Physiographic Society in Lund; UK Science and Technology Facilities Council under grant numbers ST/V001027/1 (Liverpool), ST/T004797/1 and ST/V001116/1 (Manchester); startup package from the Oregon State University College of Engineering (OSU); Office of Nuclear Regulatory Research, Nuclear Regulatory Commission under award 31310022M0019 (SJSU); U.S. Department of Energy under contract DE-AC52-07NA27344 (LLNL) and DE-AC05-00OR22725 (ORNL); Swiss National Science Foundation under grant number 200020$\_$196981; CNRS research funds (IPHC). The authors are indebted (for the use of $^{244}$Pu) to the U.S. Department of Energy, through the transplutonium element production facilities at Oak Ridge National Laboratory.

\bibliography{E116_biblio}

\end{document}


\title{Supplemental Material\\
  Towards the Discovery of New Elements:\\
  Production of Livermorium ({\it Z}\,=\,116) with $^{50}$Ti}
  
\maketitle

The Supplemental Material provides (i) detailed results and statistical assessments in the analysis of events stemming from decay chains starting with the isotope $^{290}$Lv, and (ii) details concerning the search parameters for decay chains originating from $^{291}$Lv. 

The numbering of references in this Supplemental Material corresponds to references in the main article.

\section{Details of Published $^{290}$Lv Decay Chains}
\begin{figure*}[!h]
    \includegraphics[width=180mm]{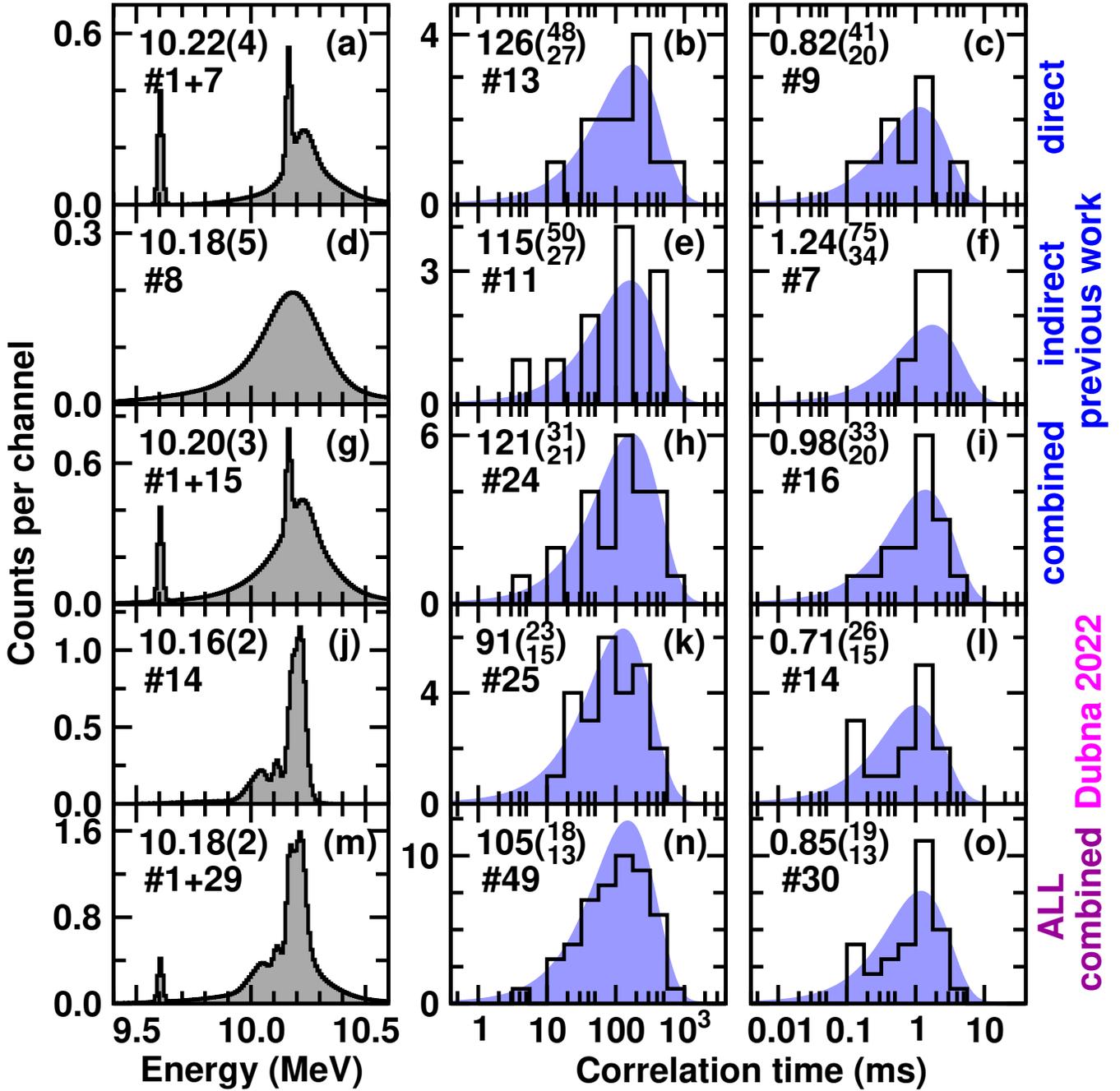}
    \caption{(Color online)
    The left column provides experimental decay-energy spectra from events
    associated with the decay step $^{286}$Fl$\,\rightarrow\,^{282}$Cn. 
    For a single entry, a Gaussian with integral one and a width compliant
    with its measured uncertainty was added into the respective spectrum. 
    The numbers at the top left of each panel in the left column are the 
    ($\alpha$-decay) energies extracted by computing the histogram mean in 
    the interval [9.9,10.5]~MeV. The middle and right columns provide 
    the correlation times for the decays of $^{286}$Fl and $^{282}$Cn, 
    respectively. Experimental data points are comprised in the histograms
    (black lines). The shaded areas (blue) provide correlation-time
    distributions expected for the corresponding half life, $T_{1/2}$,
    which are given in the top left corner of each panel. For all panels, 
    the number behind the hashtag, $\#$, indicates the number of available 
    data points. The first row, panels (a)-(c), refers to previous direct 
    production of $^{286}$Fl [14,63-65]. 
    The second row, panels (d)-(f), refers to previous indirect production 
    of $^{286}$Fl [61,62,66]. 
    The spectra in the third row, panels (g)-(i), are the sums of the
    spectra in the first and second row. The fourth row, panels (j)-(l),
    refer to the recent data obtained at JINR Dubna [67]. 
    The spectra in the fifth row, panels (m)-(o), are the sums of the
    spectra in the third and fourth row, i.e.~comprise current best values
    for the main decay characteristics of $^{286}$Fl and $^{282}$Cn prior
    to the present study. For the update {\em including} the data points
    from the present study, see Fig.~2 of the main article and Table~I of this
    Supplemental Material. For the interpretation of the peak at 9.60(1)~MeV in
    panels (a), (g), and (m), see Refs.~[14,72]. 
    }
    \label{fig:Fl286}
\end{figure*}

Decay properties such as decay energies and lifetimes, relating to various ensembles of data associated with previous experiments in the direct or indirect production of $^{290}$Lv, $^{286}$Fl, or $^{282}$Cn, and with the result of the present work (cf.~main article) included, are compiled: 
Distributions of decay energies and correlation times along with determined $E_{\alpha}$ and $T_{1/2}$ values are presented for the different ensembles in Fig.~\ref{fig:Fl286} for $^{286}$Fl and $^{282}$Cn, respectively. An overview, together with a statistical assessment of the correlation times attributed to the single decay steps relevant to the current study, is presented Table~\ref{tab:corr_times} for the ensembles of decay chains corresponding to $^{290}$Lv (cf.~main article) and Fig.~\ref{fig:Fl286}, respectively. Table~\ref{hindrance} provides a summary of aggregated experimental results concerning the decays of $^{290}$Lv, $^{286}$Fl, and $^{282}$Cn.

\section{Search Parameters for Decay Chains Originating from $^{291}$Lv}
The known decay chain originating from $^{291}$Lv and its daughters typically terminates in spontaneous fission (SF) at $^{283}$Cn or $^{279}$Ds, although it has been observed to decay via emission of $\alpha$ particles to the SF-decaying $^{267}$Rf [60-67]. 
Candidates for decay chains originating from this isotope were searched for 
by looking for correlations, required to all be detected within the same pixel of the detector, that consisted of an:
\begin{enumerate}
   \item Evaporation residue (EVR) $[10<E\,$(MeV)$\,<30]$ followed by at least one $\alpha$-like particle $[9.50<E\,$(MeV)$\,<11.25]$ within 3~s, followed by a spontaneous fission-like event with $E>120$~MeV within 25~s.
   \item EVR $[10<E\,$(MeV)$\,<30]$ followed by at least two $\alpha$-like particles $[9.00<E\,$(MeV)$\,<11.25]$ within 30~s, followed by a spontaneous fission-like event with $E>120$~MeV within 2~s. 
   \item EVR $[10<E\,$(MeV)$\,<30]$ followed by three or more $\alpha$-like particles $[8.00<E\,$(MeV)$\,<11.25]$ within 300~s, followed by a spontaneous fission-like event with $E>120$~MeV within 150~min.
\end{enumerate}
These $\alpha$-particle energy ranges were chosen to fully encompass the known energy ranges for chains terminating in SF at $^{283}$Cn, $^{279}$Ds, and $^{267}$Rf, respectively. The lifetimes were chosen to accept events within five half-lives of known decays. The efficiency for detecting a decay chain originating from $^{291}$Lv under these conditions is $\approx 85 \%$.
The probability for any one of the SF-like events to combine with random background-like events to form a chain meeting the requirements listed above is $1.4\times 10^{-3}$, $2.1\times 10^{-6}$, and $1.7\times 10^{-3}$, for conditions (1)-(3), respectively. If a chain is observed within these broad parameters, then lifetimes and decay energies must also fall within accepted windows for the known isotopes, thus reducing random rates further.

\begin{table*}[!h]
\tabcolsep12pt
\caption{
\label{tab:corr_times}
Overview of correlation time analyses of single decay steps according to Ref.~[74] 
of various ensembles of decay chains associated with previous direct ('Cn\&Fl') and indirect ('Lv\&Og') production of $^{286}$Fl, recent production of $^{286}$Fl [67], and the present events interpreted to start from $^{290}$Lv, respectively. These correspond to ensembles as displayed in the respective rows of Fig.~3 in the main article and Fig.~\ref{fig:Fl286}.}
\begin{tabular}{lcccccc}
\hline\noalign{\smallskip}
Label                                 & previous   & previous         & previous         & $^{286}$Fl  & all & including\\
                                      & (Cn\&Fl)   & (Lv\&Og)         &                  & (2022)     &     & this work\\
\hline\noalign{\smallskip}
References & 
[14,63-65]
& [61,62,66] 
&& [67] \\ 
\noalign{\smallskip}\hline\noalign{\smallskip}
$^{290}$Lv\\
No.~of chains                         &            & 16               & 16               &        & 16              & 18 \\
data points & & 11 & 11 & & 11 & 13\\
$T_{1/2}(^{290}\mathrm{Lv})$ (ms)        &            &$8.3(^{36}_{19})$   &$8.3(^{36}_{19})$   &        & 8.3($^{36}_{19}$) & 8.2($^{32}_{18}$)\\
$\sigma_{\Theta_{\rm exp}}$   &            & 2.13\footnotemark[2]& 2.13\footnotemark[2]&& 2.13\footnotemark[2]& 2.02\footnotemark[2]\\
$[\sigma_{\Theta,\rm low},\sigma_{\Theta,\rm high}]$ [74] 
                                     &            & [0.67,1.81]    & [0.67,1.81]         &        & [0.67,1.81]     & [0.72,1.77] \\
data points                          &            & 11             & 11                  &        & 11              & 12\\
$E_{\rm decay}$ (MeV)                  &           &  10.84(8)& 10.84(8) & &  10.84(8) &  10.84(7)\footnotemark[1]  \\
\noalign{\smallskip}\hline\noalign{\smallskip}
$^{286}$Fl\\
No.~of chains                         & 13          & 16              & 29               & 25                & 54 & 56 \\
data points & 13 & 11 & 24 & 25 & 49 & 51 \\
$T_{1/2}(^{286}\mathrm{Fl})$ (ms)       &$126(^{48}_{27})$&$115(^{57}_{31})$  &$121(^{31}_{21})$  &$91(^{23}_{15})$    &105($^{18}_{13}$) & 106($^{17}_{13}$) \\
$\sigma_{\Theta_{\rm exp}}$   &  0.96    &  1.44       & 1.22       &  0.97         &  1.10 &  1.11\\
$[\sigma_{\Theta,\rm low},\sigma_{\Theta,\rm high}]$ [74] 
                                     & [0.72,1.77]   & [0.67,1.81]    & [0.84,1.72]     & [0.85,1.71]      & [0.97,1.57]  & [0.98,1.58]\\
data points                          & 7 & 8 & 15 & 14 & 29 & 31 \\
$E_{\rm decay}$ (MeV)                  & 10.22(4) & 10.18(5) & 10.20(3) & 10.16(2) & 10.18(2) & 10.18(2)\footnotemark[3]  \\
\noalign{\smallskip}\hline\noalign{\smallskip}
$^{282}$Cn\\
data points & 9 & 7 & 16 & 14 & 30 & 32 \\
$T_{1/2}(^{282}\mathrm{Cn})$ (ms)       &$0.82(^{41}_{20})$&$1.24(^{75}_{34})$  &$0.98(^{33}_{20})$  & 0.71($^{26}_{15}$) & 0.85($^{19}_{13}$) & 0.88($^{19}_{13}$)\\
$\sigma_{\Theta_\mathrm{exp}}$             & 0.99    &  0.43\footnotemark[4] &  0.87       &  1.04         &  0.98 & 0.97\\
$[\sigma_{\Theta,\rm low},\sigma_{\Theta,\rm high}]$ [74] 
                                     & [0.62,1.84]   & [0.52,1.87]    & [0.77,1.75]     & [0.73,1.77]      & [0.89,1.67] & [0.90,1.66]\\
\hline\noalign{\smallskip}
\end{tabular}
\footnotemark[1]{Result from the integration of the energy spectrum in Fig.~3(a) of the main article in the interval [10.0,11.7]~MeV.} \\
\footnotemark[2]{The experimental value for $\sigma_{\Theta_{\rm exp}}$ falls outside the upper 1-$\sigma$ confidence limit [74].} \\ 
\footnotemark[3]{Result from the integration of the energy spectra in the left column of Fig.~1 in
  the interval [9.9,10.5]~MeV.} \\
\footnotemark[4]{The experimental value for $\sigma_{\Theta_{\rm exp}}$ falls outside the lower 1-$\sigma$ confidence limit [74].}\\
\end{table*}

\begin{table*}[t]
\caption{
  Summary of aggregated experimental results concerning the decays of $^{290}$Lv, $^{286}$Fl, and $^{282}$Cn.
}
\label{hindrance}
\tabcolsep4.8mm
\begin{tabular}{cccccccc}
\hline\noalign{\smallskip}
&$E_\alpha$ & $Q_\alpha$ & $T_{1/2}$ & $b_\alpha$\footnotemark[1] & $T_{1/2}(\alpha)$\footnotemark[2] & HF\footnotemark[2] & $T_{1/2}$(SF)\footnotemark[2]\\
&(MeV) & (MeV) & (ms) & & (ms) & 
& (ms)\\
\hline\noalign{\smallskip}
$^{290}$Lv & 10.84(7) & 10.99(7) & $8.2(^{32}_{18})$ & 1 & $8.2(^{32}_{18})$ & $1.4(^{15}_{\;7})$ & not applicable\\
$^{286}$Fl & 10.18(2) & 10.32(2) & $106(^{17}_{13})$ & 31/56 & $191(^{31}_{23})$& $2.2(^{7}_{5})$ & $247(^{40}_{30})$\\
& 9.57(3)\footnotemark[3] &  9.71(3) && 1/56 & $5.9(^{10}_{\;7})\times 10^3$ & $1.3(^{5}_{4})$ & \\
$^{282}$Cn & \multicolumn{2}{c}{not applicable} & $0.88(^{19}_{13})$ & 0 & \multicolumn{2}{c}{not applicable} & $0.88(^{19}_{13})$ \\
\noalign{\smallskip}\hline
\end{tabular}
\footnotemark[1]{Due to incomplete knowledge from some previous studies, the branching ratio can only be estimated.}\\
\footnotemark[2]{The uncertainties cannot account for uncertainties in 
branching ratios. See preceding note. Hindrance factors were calculated based on C.~Qi {\it et al.}, 
Phys.~Rev.~C 80, 044326 (2009). See also Refs.~[14,72].}\\
\footnotemark[3]{See corresponding note in Table II in Ref.~[14].} 
\end{table*}